\documentstyle[twoside]{article}

\catcode`\@=11
\long\def\@makefntext#1{
\protect\noindent \hbox to 3.2pt {\hskip-.9pt  
$^{{\eightrm\@thefnmark}}$\hfil}#1\hfill}		

\def\@makefnmark{\hbox to 0pt{$^{\@thefnmark}$\hss}}	
	
\def\ps@myheadings{\let\@mkboth\@gobbletwo
\def\@oddhead{\hbox{}
\rightmark\hfil\eightrm\thepage}   
\def\@oddfoot{}\def\@evenhead{\eightrm\thepage\hfil
\leftmark\hbox{}}\def\@evenfoot{}
\def\sectionmark##1{}\def\subsectionmark##1{}}

\oddsidemargin=\evensidemargin
\newcounter{sectionc}\newcounter{subsectionc}\newcounter{subsubsectionc}
\renewcommand{\section}[1] {\vspace{12pt}\addtocounter{sectionc}{1} 
\setcounter{subsectionc}{0}\setcounter{subsubsectionc}{0}\noindent 
	{\tenbf\thesectionc. #1}\par\vspace{5pt}}
\renewcommand{\subsection}[1] {\vspace{12pt}\addtocounter{subsectionc}{1} 
	\setcounter{subsubsectionc}{0}\noindent 
	{\bf\thesectionc.\thesubsectionc. {\kern1pt \bfit #1}}\par\vspace{5pt}}
\renewcommand{\subsubsection}[1] {\vspace{12pt}\addtocounter{subsubsectionc}{1}
	\noindent{\tenrm\thesectionc.\thesubsectionc.\thesubsubsectionc.
	{\kern1pt \tenit #1}}\par\vspace{5pt}}

\newcommand{\textlineskip}{\baselineskip=13pt}
\newcommand{\smalllineskip}{\baselineskip=10pt}

\def\eightcirc{
\begin{picture}(0,0)
\put(4.4,1.8){\circle{6.5}}
\end{picture}}
\def\eightcopyright{\eightcirc\kern2.7pt\hbox{\eightrm c}} 

\newcommand{\copyrightheading}[1]
	{\vspace*{-2.5cm}\smalllineskip{\center
       {\footnotesize $\eightcopyright$\, World Scientific, Mod. Phys. Lett. A 17 no. 21 (2002) 1377-1381\\ Los Alamos arXiv gr-qc/0205085 v2
       }\\
	 }}

\def\abstracts#1#2#3{{
	\centering{\begin{minipage}{4.5in}\baselineskip=10pt\footnotesize
	\parindent=0pt #1\par 
	\parindent=15pt #2\par
	\parindent=15pt #3
	\end{minipage}}\par}} 

\def\keywords#1{{
	\centering{\begin{minipage}{4.5in}\baselineskip=10pt\footnotesize
	{\footnotesize\it Keywords}\/: #1
	 \end{minipage}}\par}}

\newcommand{\bibit}{\nineit}

\renewenvironment{thebibliography}[1]
	{\frenchspacing
	 \ninerm\baselineskip=11pt
	 \begin{list}{\arabic{enumi}.}
        {\usecounter{enumi}\setlength{\parsep}{0pt}     
	 \setlength{\leftmargin 12.7pt}{\rightmargin 0pt} 
         \setlength{\itemsep}{0pt} \settowidth
	{\labelwidth}{#1.}\sloppy}}{\end{list}}

\newcounter{itemlistc}
\newcounter{romanlistc}
\newcounter{alphlistc}
\newcounter{arabiclistc}

\def\@citex[#1]#2{\if@filesw\immediate\write\@auxout
	{\string\citation{#2}}\fi
\def\@citea{}\@cite{\@for\@citeb:=#2\do
	{\@citea\def\@citea{,}\@ifundefined
	{b@\@citeb}{{\bf ?}\@warning
	{Citation `\@citeb' on page \thepage \space undefined}}
	{\csname b@\@citeb\endcsname}}}{#1}}

\newif\if@cghi
\def\cite{\@cghitrue\@ifnextchar [{\@tempswatrue
	\@citex}{\@tempswafalse\@citex[]}}
\def\citelow{\@cghifalse\@ifnextchar [{\@tempswatrue
	\@citex}{\@tempswafalse\@citex[]}}
\def\@cite#1#2{{$\null^{#1}$\if@tempswa\typeout
	{IJCGA warning: optional citation argument 
	ignored: `#2'} \fi}}

\def\@refcitex[#1]#2{\if@filesw\immediate\write\@auxout
	{\string\citation{#2}}\fi
\def\@citea{}\@refcite{\@for\@citeb:=#2\do
	{\@citea\def\@citea{, }\@ifundefined
	{b@\@citeb}{{\bf ?}\@warning
	{Citation `\@citeb' on page \thepage \space undefined}}
	\hbox{\csname b@\@citeb\endcsname}}}{#1}}

\def\@refcite#1#2{{#1\if@tempswa\typeout
        {IJCGA warning: optional citation argument
	ignored: `#2'} \fi}}

\def\refcite{\@ifnextchar[{\@tempswatrue
	\@refcitex}{\@tempswafalse\@refcitex[]}}

\def\pmb#1{\setbox0=\hbox{#1}
	\kern-.025em\copy0\kern-\wd0
	\kern.05em\copy0\kern-\wd0
	\kern-.025em\raise.0433em\box0}

\def\fnt#1#2{\footnotetext{\kern-.3em
	{$^{\mbox{\scriptsize #1}}$}{#2}}}

\font\tenrm=cmr10
\font\tenit=cmti10 
\font\tenbf=cmbx10
\font\bfit=cmbxti10 at 10pt
\font\ninerm=cmr9
\font\nineit=cmti9

\font\eightrm=cmr8

\def\qed{\hbox{${\vcenter{\vbox{			
   \hrule height 0.4pt\hbox{\vrule width 0.4pt height 6pt
   \kern5pt\vrule width 0.4pt}\hrule height 0.4pt}}}$}}

\begin{document}

\normalsize\textlineskip
\thispagestyle{empty}
\setcounter{page}{1}

\copyrightheading{}                     

\vspace*{0.88truein}

\centerline{\bf ON ARITHMETIC DETECTION OF GREY PULSES}
\centerline{\bf WITH APPLICATION TO HAWKING RADIATION }
\vspace*{0.23truein}
\vspace*{0.37truein}
\centerline{\footnotesize  HARET C. ROSU}
\centerline{\footnotesize hcr@ipicyt.edu.mx}
\vspace*{0.015truein}
\centerline{\footnotesize\it Dept. of Applied Mathematics, IPICyT, 
Apdo Postal 3-74 Tangamanga, San Luis Potos\'{\i}, MEXICO}
\vspace*{10pt}
\centerline{\footnotesize  MICHEL PLANAT}
\centerline{\footnotesize planat@lpmo.edu}
\vspace*{0.015truein}
\centerline{\footnotesize\it  Laboratoire de Physique et M\'etrologie des Oscillateurs du CNRS,
25044 Besan\c{c}on Cedex, FRANCE}
\baselineskip=10pt
\vspace*{0.225truein}

\vspace*{0.21truein}
\abstracts{
Micron-sized black holes do not necessarily have a constant horizon temperature distribution.
The black hole remote-sensing problem means to find out the `surface' temperature distribution of a
small black hole from the spectral
measurement of its (Hawking) grey pulse.  This problem has been previously considered by Rosu, who used Chen's 
modified M\"obius inverse transform. Here, we hint on a Ramanujan generalization of Chen's modified
M\"obius inverse transform that may be considered as a special wavelet 
processing of the remote-sensed grey signal coming from a black hole or any other distant grey source.\\
}{}{}

\vspace*{10pt}

\textlineskip                  
\vspace*{12pt}                 

\vspace*{1pt}\textlineskip	
\vspace*{-0.5pt}


\noindent

\noindent


\section{INTRODUCTION}

\noindent
There may exist black holes in the micron size range carrying on some external
distribution of matter. Theoretical examples are 
(i) down-scaled {\em Weyl black holes},\cite{w} for which the metric potentials are solutions of polar 
Laplace equations
(ii) {\em black holes with Einstein shells},\cite{e}
(iii) {\em primordial black holes} (PBH) and/or any
{\em mini black hole} hovering through the universe and carrying on some
matter distributions, 
(iv) {\em hairy black holes},\cite{bow} with additional conserved quantum numbers beyond those allowed 
by the classical no hair theorems and {\em dirty black holes} in the sense of Visser,\cite{v} i.e., black
holes in interaction with various classical fields, for which the Hawking
temperature appears to be supressed relative to the vacuum black holes of
equal area.
In some cases, for small enough black holes, the external distribution of
matter can be of such a kind as to disturb only slightly the pure horizon
Hawking radiation and consequently from the praxis standpoint we have a
grey-body radiation problem. 

Hawking radiation by itself is distorted with respect to a pure black-body spectrum,
especially
in the low frequency regime due to a grey-body factor usually identified
with the square of the absorption amplitude for the mode.\cite{be}
A useful work on the nature of the grey body problem for black holes has been written by Schiffer.\cite{s}

In this letter, we first review the grey body inverse problem and the modified M\"obius inverse transform (Chen's transform)
in sections 2 and 3, respectively.\cite{ro} In section 4, we hint on a possible Ramanujan extension of Chen's transform
with possible application to small black holes and the cosmological background radiation.

\bigskip

\section{INVERSE GREY-BODY PROBLEM}

\noindent

\noindent
Planck's law provides
the analytical formula for the emitted power
spectrum from black body sources.  In laboratory physics the emitted power spectrum is also called
spectral brightness, or spectral radiance of the black body radiation.
The latter notion is used in radiometry to characterize the 
spectral properties of the source as a function of position and direction from the source. For point, i.e.,
far away, grey sources the total radiated power spectrum, also called radiant
spectral intensity is
\begin{equation} \label{w1} W(\nu)\sim \int _{0}^{\infty} A(T)B(\nu , T)dT~,   \end{equation}
where 

$A(T)$ is the area temperature distribution of the grey body, 

$B(\nu , T)$ is the Boltzmann-Planck occupation factor. 

\noindent
Finding out $A(T)$ at given $W(\nu)$ is known as 
the inverse grey-body problem.\cite{l}
$W(\nu)$ may be known either experimentally or within some theoretical
model. This inverse problem was solved in principle by Bojarski,\cite{bo}
by means of a
thermodynamic coldness parameter $u=h/kT$, and an area coldness distribution
$a(u)$, as more convenient variables than $T$ and $A(T)$ to get an inverse Laplace transform of
the total radiated power. The coldness distribution is obtained as an expansion
in this Laplace transform.
Explicitly, the total grey power spectrum is rewritten as an integral over the coldness variable
\begin{equation} \label{w2} 
W(\nu)=\frac{2h\nu ^{3}}{c^{2}}\int _{0}^{\infty}\frac{a(u)}{\exp(u\nu)-1}du
 \end{equation}
and furthermore as
\begin{equation} \label{w3} 
W(\nu) =\frac{2h\nu ^{3}}{c^{2}}\int _{0}^{\infty}\exp(-u\nu)
\Big[\sum _{n=1}^{\infty}(1/n)a(u/n)\Big]du~.   
\end{equation}
Therefore the sum under the integral that we shall denote by $f(u)$ is
the Laplace transform of $g(\nu)=\frac{c^{2}}{2h\nu ^{3}}W(\nu)$ and the inverse 
Laplace transform of $g$ will provide the sought coldness distribution.

Despite the formal mathematical solution the inverse grey body problem  is unstable for most numerical
implementations, i.e., it belongs to the broad class of ill-posed inverse problems.\cite{x}

\newpage

\section{MODIFIED M\"OBIUS TRANSFORM (MMT)} 

\noindent
Chen,\cite{c} obtained $a(u)$ by
means of the so-called modified M\"obius transform ( MMT) of $f(u)$
\begin{equation} \label{a}
a(u)=\sum _{n=1}^{\infty}\frac{\mu(n)}{n}f(u/n)~.
\end{equation}

To understand Eq.~(\ref{a})  we recall a few basic results from the theory of numbers.\cite{ba}
The M\"obius expansion refers to special sums over prime factor divisors,
(d-sums)  running over all the prime factors of $n$, 1 and $n$ included, of any
 function $f(n)$ defined on the positive integers
\begin{equation} \label{f1}
S_{f}(n)=\sum_{d|n}^{n} f(d)~.
\end{equation}
The remarkable fact in this case is that the {\em last} term of the sum can be
written in turn as a sum over the  $S_f$ arithmetical functions. The latter sum is called the inverse M\"obius
transform (or the M\"obius d-sum) of $f$
\begin{equation}\label{f2}
f(n)=\sum _{d|n}^{n} \mu (d)S_f(n/d)~,
\end{equation}
in which the d-sum $S_f(n)$ becomes the {\em first} term of the M\"obius
d-sum, and where $\mu(d)$ is the famous M\"obius function.
Since at the left hand side of (\ref{f2})  one has only a term of a d-sum whereas on the right hand side there is a sum of d-sums
there is clear overcounting, unless the M\"obius function is sometimes either
naught or negative. The partition of the prime factors of $n$ implied by the
M\"obius function is such that, by definition, $\mu(1)$ is 1, $\mu(n)$ is
$(-1)^{r}$ if $n$ includes $r$ distinct prime factors, and $\mu(n)$ is naught
in all the other cases. In particular, all the squares have no contribution to
the inverse M\"obius transforms. That is why the integers selected by the
M\"obius function are also called square-free integers.

Chen's MMT means to apply such an inversion of finite sums to infinite
summations, and to ordinary functions of real continuous variable(s). MMT
means that if
\begin{equation} \label{f3}
y_{1}(x)= \sum_{n=1}^{\infty} y_{2}(x/n)~,
\end{equation}
then
\begin{equation}\label{f4}
y_{2}(x)= \sum_{n=1}^{\infty} \mu (n) y_{1}(x/n) ~. 
\end{equation}
For the inverse grey-body problem, $y_{1}(u)=uf(u)$ and $y_{2}(u)=ua(u)$.
So, one can get the coldness distribution by multiplying the Laplace transform
of the total power spectrum by the coldness parameter, and then applying
the MMT.

\newpage

\section{RAMANUJAN GENERALIZATION OF CHEN'S TRANSFORM}

\noindent
Ramanujan sums are well known in number theory  but only recently some physical applications
have been suggested.\cite{p} They are of the form
\begin{equation}\label{ram}
c_q(m)= \sum_{p=1}^{q} \cos(2\pi mp/q )~,
\end{equation}
with irreducible fractions $p/q$. 
The sums are quasiperiodic in $m$ and aperiodic in the denominator $q$. They are a generalization 
of the M\"obius function since $c_q(m)=\mu (q)$ whenever $q$ and $m$ are coprimes.
The MMT for black holes in the Ramanujan notation will be
\begin{equation}\label{f4}
[ua(u)]_{(1)}= \sum_{q=1}^{\infty} c_q (1) uf(u/q)~,  
\end{equation}
but the point is that many other Ramanujan inverse transforms can be introduced through
\begin{equation}\label{f4}
[ua(u)]_{(i)}= \sum_{q=1}^{\infty} c_q (i) uf(u/q) ~, 
\end{equation}
for integers $i>1$ and $(q,i)=1$. One can see that a sequence of two-dimensional analyses $(i,q)$ of the signal are available in this 
approach similarly to the time-frequency analysis that is so characteristic to wavelets.


\noindent
As an example, for micron-sized Schwarzschild black holes ($M\sim 10^{24}$ g),
no known massive particles are thermally emitted, and according to the
calculations
of Page,\cite{pa} about 16\% of the Hawking flux goes into photons, the rest
being neutrino emission. Let us consider these black holes as grey objects of the following two classes
(i) by their own,\cite{be}
and (ii) of the Weyl type. The coldness
parameter will be in the first case
$u_{S}=\frac{1}{\nu}\ln\left(1+
\frac{e^{\beta _{h}\hbar \omega}-1}{\Gamma (\omega)}\right)$, where
$\beta _{h}$ is the horizon inverse temperature parameter,
and $\Gamma (\omega)$ is
the penetration factor of the curvature and angular momentum barrier
around the black hole,\cite{be} whereas in the latter
case $u_{S}=h/kT_{d}$, where $T_{d}$ can be considered as an
effective horizon temperature of the distorted black holes
$T_{d}=(8\pi M)^{-1}\exp (2\cal U)$, where 
${\cal U}$ is given in the work of Geroch and Hartle.
With the coldness parameter at hand  one can apply the aforementioned number
methods for the black hole emissivity problem.

\noindent
Before closing, we mention that an alternative viewpoint on MMT developed by Hughes {\em et al},\cite{hu}
in terms of Mellin transform and Riemann's $\zeta$ function is also extremely interesting if one takes
into account that the Bekenstein-Mukhanov spectrum,\cite{bm} 
could be considered as the eigenvalue problem of relativistic Schroedinger equations in finite differences.\cite{ber}


\noindent
Thus, it appears that a richer information content of the spectrum of important astrophysical 
signals that may reveal hidden discrete features can be obtained by extensive use of 
number theoretical techniques. The cosmological background radiation signal can be studied by the same approach and with the
same aim. There are claims by Hogan,\cite{ho} that inflationary perturbations display discreteness not predicted by the standard field theory
and that this discreteness may be observable in cosmic background anisotropy.


\end{document}